\documentclass{article}
\usepackage{spconf,amsmath,graphicx}
\usepackage{booktabs}
\usepackage[%
    colorlinks=false,
    pdfborder={0 0 0},
    linkcolor=black
]{hyperref}

\title{The HCCL-DKU System for fake audio generation task of the 2022 ICASSP ADD Challenge}
\name{Ziyi Chen$^{1,2}$, Hua Hua$^{1,2}$, Yuxiang Zhang$^{1, 2}$, Ming Li$^{3}$, Pengyuan Zhang$^{1,2}$}
\address{ $^1$Key Laboratory of Speech Acoustics \& Content Understanding, Institute of Acoustics, CAS, China\\
	$^2$University of Chinese Academy of Sciences, Beijing, China\\
	$^3$Data Science Research Center, Duke Kunshan University, Kunshan, China\\
	\{chenziyi, zhangpengyuan\}@hccl.ioa.ac.cn, ming.li369@duke.edu}
\begin{document}
%
\maketitle
\begin{abstract}

The voice conversion task is to modify the speaker identity of continuous speech while preserving the linguistic content. Generally, the naturalness and similarity are two main metrics for evaluating the conversion quality, which has been improved significantly in recent years. This paper presents the HCCL-DKU entry for the fake audio generation task of the 2022 ICASSP ADD challenge. We propose a novel ppg-based voice conversion model that adopts a fully end-to-end structure. Experimental results show that the proposed method outperforms other conversion models, including Tacotron-based and Fastspeech-based models, on conversion quality and spoofing performance against anti-spoofing systems. In addition, we investigate several post-processing methods for better spoofing power. Finally, we achieve second place with a deception success rate of 0.916 in the ADD challenge. 

\end{abstract}
\section{Introduction}
\label{sec:intro}
The first Audio Deep Synthesis Detection Challenge spurs researchers worldwide into building innovative techniques that can further accelerate and foster research on detecting deep fake audios. This challenge includes three fake audio detection tasks and a contrasting task that generates fake audio. In this paper, we present our proposed system for the fake audio generation task (Track 3.1). Both multi-speaker text-to-speech (TTS) and voice conversion techniques can be used to generate synthesized audios for Track 3.1. Nevertheless, since the voice conversion attains better spoofing performance against detection systems, we choose to use voice conversion in this challenge.

 Voice conversion (VC) is a technique that converts a source speaker's voice to a target speaker's voice without changing the linguistic information \cite{sisman2020overview}. With the development of deep learning, there are plenty of researches on deep learning based voice conversion. Generative adversarial network and its variants, such as StarGAN-VC \cite{kameoka2018stargan} and CycleGAN-VC \cite{kaneko2018cyclegan} use generator or conditional generator to transform the source speaker's features to target speakers' features directly. Autoencoder based model, such as, AutoVC\cite{qian2019autovc}, VQVC\cite{wu20p_interspeech} and so on. However, benefiting from acoustic features extracted by ASR models can keep the linguistic information while removing most speaker information. So ppg-based and cascading ASR and TTS model show excellent performance both in conversion quality and stability.

Nevertheless, the traditional cascading pipeline is rather complicated, and there are cascaded errors and over-smoothing issues across speech modules. Therefore, we proposed a fully end-to-end ppg-based VC model inspired by VITS \cite{kim2021conditional}. Specifically, our proposed model incorporates a conformer encoder from a pre-trained ASR model and a few transformer blocks on mel spectrograms, while we use a posterior encoder upon linear spectrograms. Outputs from those two modules are constraint to be from the same distribution. Then reparameterization process and the HiFiGAN decoder are followed to convert the hidden features to the target waveform. Moreover, we explore several physical spoofing strategies to improve the performance of attacks against anti-spoofing systems. 

 The rest of the paper is organized as follows: Section 2 describes our proposed method thoroughly. Experimental results are shown in Section 3, while the conclusion is drawn in Section 4.

\begin{figure*}[]

\begin{center}

\includegraphics[width=15cm]{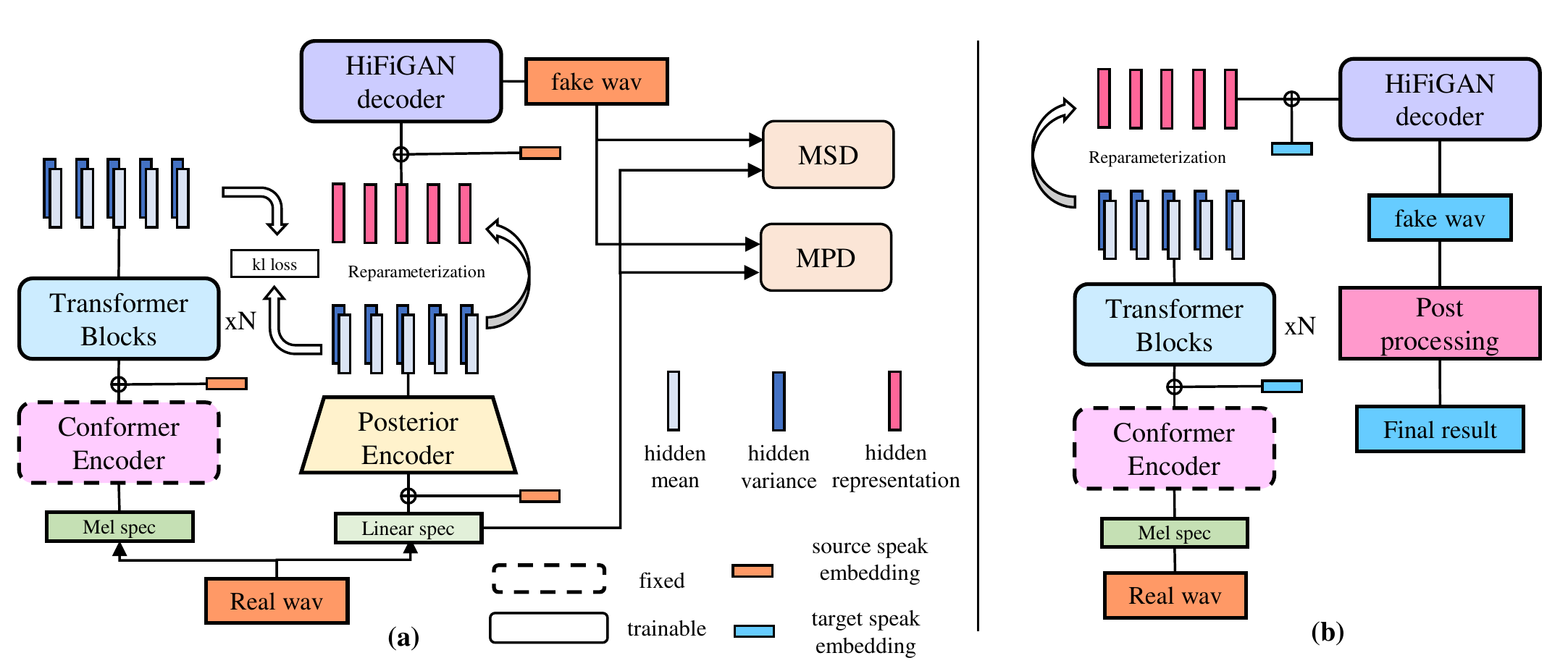}

\end{center}

\caption{Proposed method structure: (a) training procedure (b) inference procedure}

\label{vitsvc}

\end{figure*}

\section{Methods}
\label{sec:format}

This section describes the details of our proposed voice conversion architecture along with the training procedures. Basically, we utilize an encode from a pre-trained hybrid CTC/Attention transformer-based ASR model to extract Phonetic PosteriorGrams (PPGs). Then PPGs are converted to raw waveform by a fully end-to-end model.

\subsection{Model structure}
\label{subsec:modelstructure}
Fig.\ref{vitsvc} showes the overall pipeline of the proposed model. The model can be divided into five components and they are introduced in detail in the following sections.

\subsubsection{Conformer Encoder}
\label{subsec:conformer}
Conformer \cite{gulati20_interspeech} is a convolution-augmented transformer for ASR. It is a stack of two feed-forward modules, a self-attention module and a convolution module.The conformer encoder takes the mel spectrum $m$ as input to generate the hidden linguistic embedding $g$. 

For convenience, we adopt the implementation of Wenet\footnote{https://github.com/wenet-e2e/wenet}\cite{yao2021wenet} and its pretrained model on WenetSpeech\footnote{https://wenet-e2e.github.io/WenetSpeech/}\cite{zhang2021wenetspeech} as the conformer encoder of the proposed implementation.

\subsubsection{Transformer blocks}
\label{subsec:transformer}

We adopt Feed-Forward Transformer as the encoder in conversion model. It includes a feed-forward structure based on Multi-Head Self-Attention and 1D convolutions and this modules was firstly proposed in FastSpeech\cite{ren2019fastspeech}, which has been one of the most popular TTS framework. The input to the transformer blocks, which is a concatenation of hidden linguistic information $g$ and a speaker embedding $s$, is converted to the mean $q_{mean}$ and log variance $q_{logvar}$  of a gaussian distribution by this module.

\subsubsection{Posterior Encoder}
\label{subsec:posterior}

For the posterior encoder, the non-causal WaveNet\cite{oord2016wavenet} residual blocks used in WaveGlow\cite{prenger2019waveglow} were adopted. The WaveNet residual block consists of dilated convolutions with skip connection and a gated activation. And for multi-speaker task, we add speaker embedding in residual blocks by global conditioning. The posterior encoder take linear spectrum $x_{linear}$ and speaker embedding $s$ as inputs to produce the mean $p_{mean}$ and log variance $p_{logvar}$ of a gaussian distribution.

\subsubsection{HiFiGAN decoder}
\label{hifigan}

The decoder is almost essentially the HiFi-GAN generator\cite{kong2020hifi}. It is a stack of convolution blocks, which include transpose convolution and multi-receptive field fusion model(MRF). MRF is composed of residual blocks with different receptive field size. To avoid possible checkerboard artifacts caused by transpose convolution, we use temporal nearest interpolation layer followed by 1-D convolution layer as the upsampling layer\cite{pons2021upsampling}. The input of HiFiGAN decoder includes hidden representation $z$ and speaker embedding $embed$. And $z$ can be represented in Eq.\ref{eq2} during training.

\begin{equation}
    z = p_{mean} + e^{0.5*p_{logvar}} * \mathcal{N}(0,1)
    \label{eq2}
\end{equation}

The HiFiGAN decoder takes hidden representation $z$ and speaker embedding $s$ as input to get generated $w_{g}$.

\subsubsection{Discriminator}
\label{subsec:disc}
The discriminators includes multi-period discriminator(MPD) and multi-scale discriminator(MPD).  MPD is a mixture of window-based sub-discriminators, each of which operates on different periodic patterns of waveform. MSD directly operate on time domain of different scales.

\subsection{KL divergence}

Traditional TTS and VC methods cascade acoustic model and vocoder by using mel spectrum as the intermediate feature, which may cause cascaded error and over-smoothing due to mean squared error  optimization. Due to the transformer block and posterior encoder predict the mean and variance of gaussian distributions, we can use KL divergence to measure the distance of two distributions. KL divergence of two univariate gaussians can be represented in Eq.\ref{eqkl}.

\begin{equation} \begin{aligned}
    KL(p, q) &= L_{kl} \\
             &= log(\frac{\sigma_q}{\sigma_p}) + \frac{\sigma_p^2 + (\mu_p -\mu_q)^2}{2\sigma_q^2} - \frac{1}{2}
    \label{eqkl}
\end{aligned}\end{equation}

where $\sigma_p=e^{0.5p_{logvar}}$, $\sigma_q=e^{0.5q_{logvar}}$, $\mu_p=p_{mean}$ and $\mu_q=q_{mean}$ in our proposed method.

\subsection{Adversarial Training}

As with GAN based vocoders, we also add a discriminator $D$ that distinguishes audio $w_g$ generated by generator $G$ and ground truth audio $w_r$. We use least-squares as the adversarial loss for its stablity and feature-matching loss for training the generator.  

\begin{equation}\begin{aligned}
    & L_{adv}(D) = E_{(w_f,w_r)}\Big[(D(w_r) - 1)^2 + D(G(w_f))^2\Big] \\
    & L_{adv}(G) = E_{w_f}\Big[(D(G(w_f)-1)^2\Big] \\
    & L_{fm}(G) = E_{(w_f,w_r)}\Big[\sum_{T}^{l=1}\frac{1}{N_l}\Vert D^l(y) - D^l(G(z))\Vert_1\Big]
\end{aligned}\end{equation}

where $T$ denotes the number of sub-discriminators and $D^l$ denotes the feature map of the $l$-th layer of sub-discriminator with $N_l$ number of features.
\subsection{Training and inferencing}
\label{subsec:traindetail}
Due to GPU memory limit, we randomly select segments from hidden representations and the corresponding waveform for training the HiFiGAN decoder. We use $L_2$ loss between the ground truth and generated mel-spectrogram as reconstruction loss.

\begin{equation}
    L_{recon} = \Vert mel(w_g) - mel(w_f) \Vert_2
\end{equation}

Combining conditional VAE and GAN training, the total loss for training the generator can be expressed as follows:

\begin{equation}
    L_{total} = L_{recon} + L_{kl} + L_{adv}(G) + L_{fm}(G)
\end{equation}

In inferencing stage, the posterior encoder is not required and also the speaker embedding is set to  the target speaker embedding. 

\section{Experiments}
\label{sec:typestyle}

\subsection{Experimental setup}
     All our experiments are conducted on open source mandarin datasets, including the 85h multi-speaker dataset AISHELL3\cite{shi2020aishell}, an 12h female dataset from  databaker\footnote{https://www.data-baker.com/\#/data/index/source}, and the 5h male and 5h female M2VoC dev set\footnote{http://challenge.ai.iqiyi.com/M2VoC}. We convert the audio from 48KHz into 24KHz for all generation experiments and 16KHz for all evaluations.
\subsection{Comparison of VC methods}
\begin{figure}[htbp]
    
    \begin{center}
    
    \includegraphics[width=7cm]{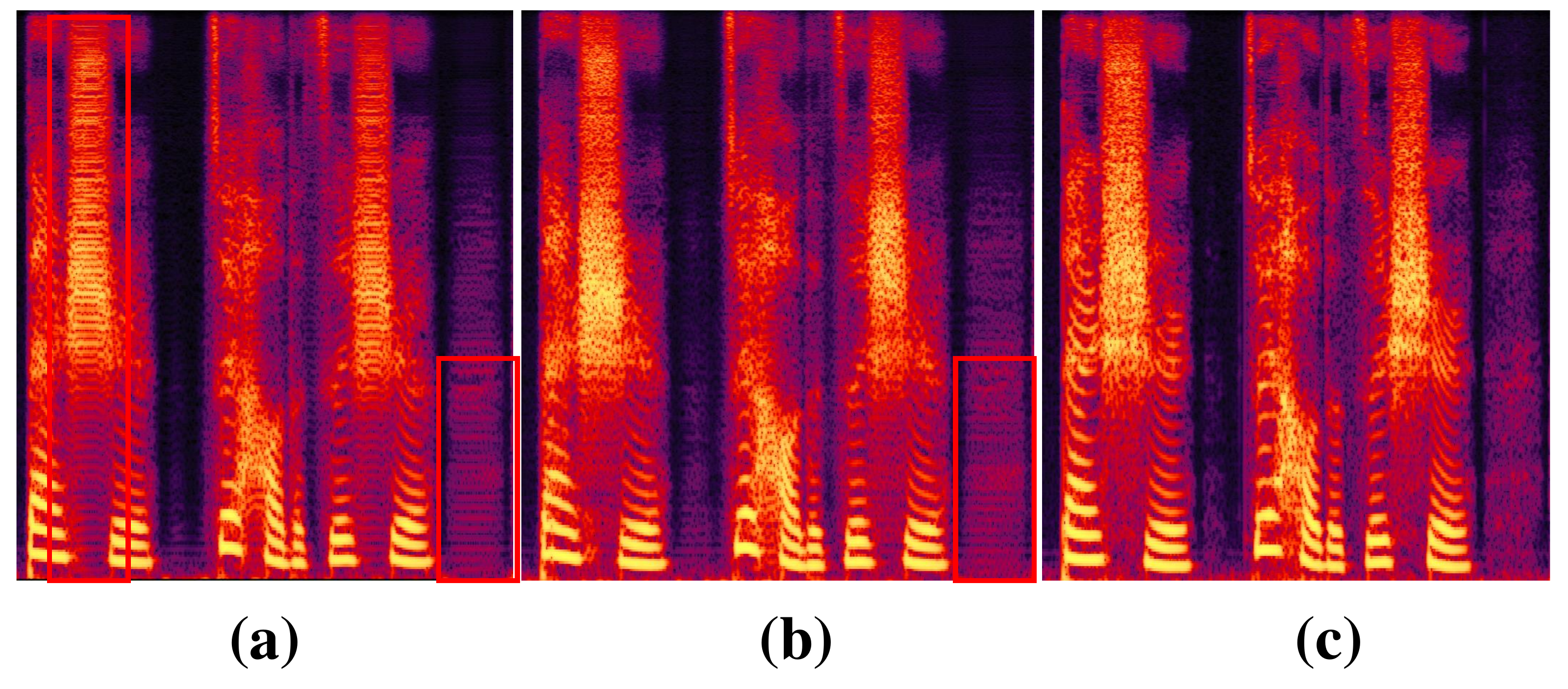}
    
    \end{center}
    
    \caption{Converted results:(a) ppg-fast (b) ppg-taco (c) proposed method}
    
    \label{spectrum}
    \end{figure}

    We conduct experiments of different vc methods with databaker and M2VoC dev set. To make it fair, all vc methods share the same conformer encoder. We modified tacotron2\cite{shen2018natural} and fastspeech as the acoustic model respectively, which we call \emph{ppg-taco} and \emph{ppg-fast} below. They share the same HiFi-GAN vocoder, which has the same structure and parameters with HiFiGAN decoder in the proposed
    method. The conversion results are shown in Fig.\ref{spectrum}, from which can be observed that there are harmonic structures in breath and unvoiced segments of \emph{ppg-taco} and \emph{ppg-fast}, which may be detected by anti-spoofing systems easily. 
 
    \begin{table*}[htbp]
        \centering
        \caption{EER of VC methods on anti-spoofing systems}
        \label{eervc1}
        \begin{tabular}{ccccccc}
        \toprule[1.5pt]
        VC method & AASIST$\uparrow$ & FFT\_SENet$\uparrow$ & MSTST\_LCNN$\uparrow$ & SILENCE$\uparrow$ & Average$\uparrow$  \\ 
        \midrule
        ppg-fast      & 41.2\%           & 46.8\%            & 35.5\%          & \textbf{60.25}\%   & 45.9\% \\
        ppg-taco      & 37.5\%           & \textbf{48.5}\%   & 31.2\%          & 47.2\%             & 41.1\% \\
        proposed      & \textbf{62.3}\%  & 39.3\%           & \textbf{51.6}\%  & 46.25\%            & \textbf{49.8}\% \\
        \bottomrule[1.5pt]
        \end{tabular}
    \end{table*}

    \begin{table*}[htbp]
        \centering
        \caption{EER of post-processing in VC and TTS methods on anti-spoofing systems}
        \label{anti}
        \begin{tabular}{ccccccc}
        \toprule[1.5pt]
        Spoof Method & Post-processing             & AASIST$\uparrow$          & FFT\_SENET$\uparrow$       & MSTFT$\uparrow$           & SILENCE$\uparrow$          & AVERAGE$\uparrow$          \\
        \midrule
        vits\_tts     & -                   & 70.0\%          & 69.5\%          & 73.4\%          & 37.2\%          & 62.5\%          \\
        proposed\_vc      & -                   & \textbf{84.3}\%          & 72.6\%          & 85.9\%          & 36.4\%          & 69.8\%          \\
        vits\_tts     & silence replacement               & 69.5\%          & 70.5\%          & 77.9\%          & 36.8\%          & 63.7\%          \\
        proposed\_vc      & silence replacement               & 84.0\% & \textbf{82.2\%} & \textbf{88.0\%} & 53.2\%          & \textbf{76.9\%} \\
        proposed\_vc      & global noise(SNR=40) & 83.8\%          & 78.3\%          & 87.9\%          & \textbf{56.8\%} & 76.7\%          \\
        proposed\_vc      & global noise(SNR=50) & 84.0\% & 81.0\%          & \textbf{88.0\%} & 53.5\%          & 76.6\%          \\
        \bottomrule[1.5pt]
        \end{tabular}
    \end{table*}

    \begin{table}[htbp]
        \centering
        \caption{Speech naturalness and Speaker similarity}
        \label{mos}
        \begin{tabular}{ccc}
        \toprule[1.5pt]
        Method      & Speech naturalness$\uparrow$ & Speaker similarity$\uparrow$ \\
        \midrule
        ground truth & 4.62$\pm$0.14           & 4.75$\pm$0.16           \\
        ppg-fast    & 3.46$\pm$0.19           & 3.69$\pm$0.20           \\
        ppg-taco    & 3.87$\pm$0.21           & 3.91$\pm$0.17           \\
        proposed    & \textbf{4.01$\pm$0.20}           & \textbf{3.98$\pm$0.19}      \\
        \bottomrule[1.5pt]
    \end{tabular}
    \end{table}
    
    AASIST\cite{jung2021aasist}, FFT\_SENet\cite{zhang21da_interspeech}, MSTST\_LCNN\cite{tomilov2021stc} and SILENCE were used to test our pre-hypothesis. 800 fake samples and 400 real samples from databaker were used as trials for test anti-spoofing EER. The results are shown in Table. \ref{eervc1}. The proposed method is obviously better than other two baseline methods. 
    
     Subjective evaluation on speech naturalness and speaker similarity of converted speech are conducted. There are ten speakers in the survey and are asked to give a 5-scale opinion score on both speaker similarity and naturalness. The subjective evaluation results are shown in Table. \ref{mos}. Generated audios can be found in the demo page\footnote{https://miracyan.github.io/2022hccldkuadd/}.

\subsection{Post-processing}

    To further enhance the attacking capability of speech produced by our proposed method, we consider adopting countermeasures against silence based anti-spoofing systems. We experimented two methods of appending silence to our generated speech. 
    
    The first one is silence replacement. Using a non-neural-network VAD system that quickly finds silence segments in fake speech, we randomly selected from our extracted real silence segments and crop them to the same length as the fake silence segments. Then we replaced the silence segments in the fake speech with the cropped real silence segments. 
    
    The second one is global noise. We randomly select multiple real silence segments and then normalize their amplitudes to average level. After that, we connect them together through the strategy of parabolic cross-fading and finally directly superimpose them to our fake speech like additive noise.
    
    We compared post-processing on audio generated from proposed method and VITS, which is trained on AISHELL3. In Table.\ref{eervc1}, the proposed method performs worse in FFT\_SENet and SILENCE worse, while in Table.\ref{anti}, the proposed method with post-processing improves a lot in FFT\_SENet and SILENCE, which verifies the validity of the proposed post-processing methods. 

\subsection{Evaluations}

Track 3.1 requires participants to generate attack samples with respect to the given text and speaker identities. For voice conversion, this task lacks source audio. As our proposed method is an any-to-many voice conversion system, we apply VITS trained on databaker to get source audios of the given text, then convert them to given speaker identities. Then post-processing of silence replacement was conducted on generated samples.

\begin{figure}[htbp]
    \label{dsrrank}
    \begin{center}

    \includegraphics[width=7cm]{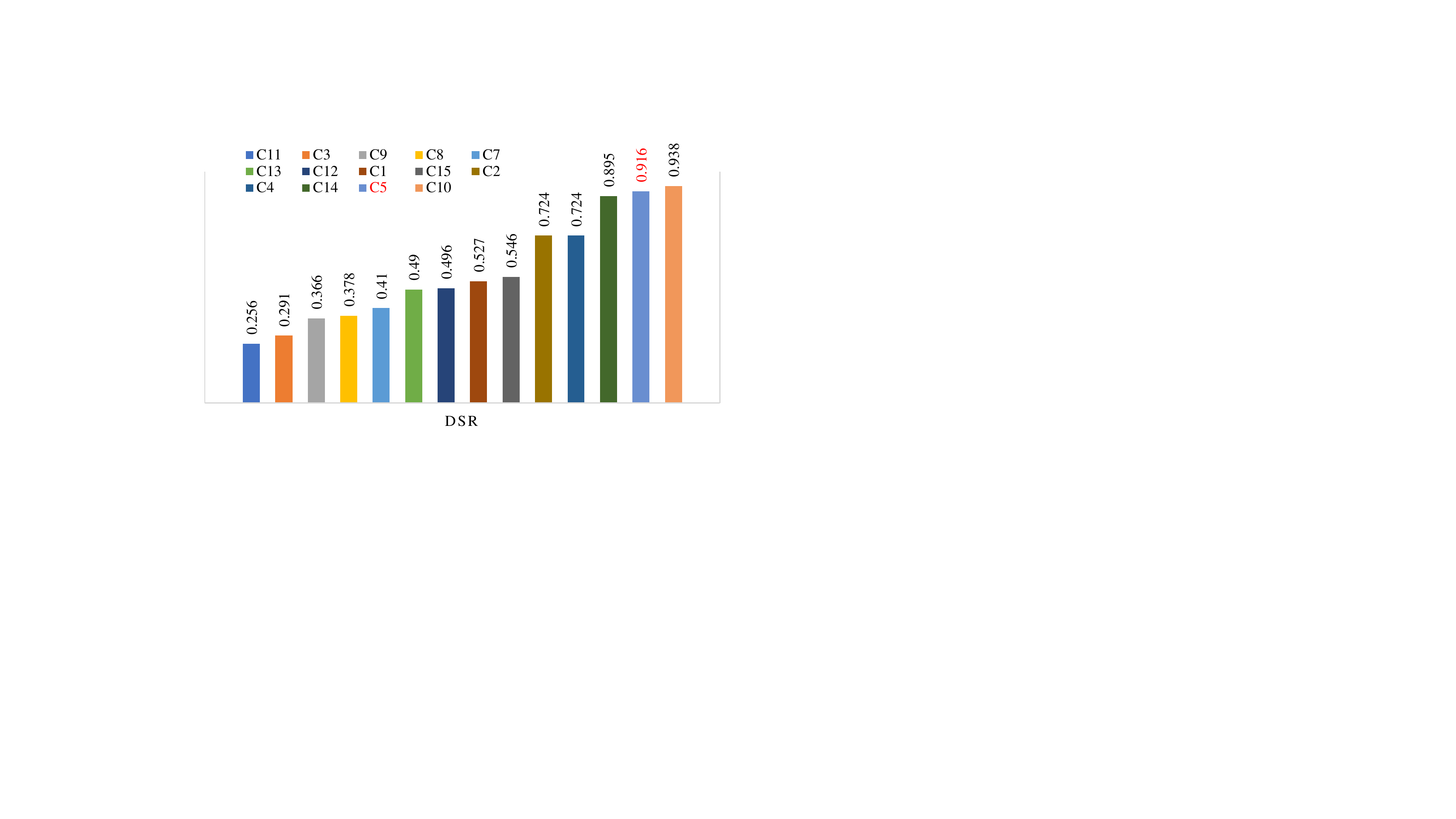}
    
    \end{center}
    
    \caption{Deception success rate(DSR) team rank}
    
    \label{table}

\end{figure}

The deception success rate of all teams is shown in Fig.3. We achieve the second place, which verifies the effectiveness of our proposed method and post-processing technique.

\section{Conclusion}
\label{sec:majhead}
In this paper, we present a novel e2e structure for any-to-many voice conversion, which performs better than baselines in terms of voice conversion quality and deception success rate. And we also proposed a simple but efficient post-processing method. 
\vfill\pagebreak
\bibliographystyle{IEEEbib}
\bibliography{strings,refs}

\end{document}